\documentclass[aps,pre,preprint,groupedaddress,amsmath,amssymb,showpacs]{revtex4}

\usepackage{graphicx}
\usepackage{bm}

\def\w{\omega}
\def\p{\phi}
\def\cx{\cos{\theta}}

\def\t{\theta}
\def\D{\Delta}

\def\pa{\p\bm{(}A\bm{)}}
\def\px{\p\bm{(}x\bm{)}}
\def\Rx{R\bm{(}x\bm{)}}
\def\Rt{R\bm{(}\theta \bm{)}}

\begin{document}

\title{Periods of relativistic oscillators with even polynomial potentials}
\author{Mikhail P. Solon}
\email{mpsolon@up.edu.ph}
\affiliation{National Institute of Physics, University of the Philippines, Diliman, Quezon City, Philippines}
\author{J.~P.~Esguerra}
\email{pesguerra@nip.upd.edu.ph}
\affiliation{National Institute of Physics, University of the Philippines, Diliman, Quezon City, Philippines}
\date[]{}
\preprint{}

\begin{abstract}
The authors modify a non-perturbative approach based on the Principle of Minimal Sensitivity to calculate the periods of relativistic oscillators with even polynomial potentials. The optimization of the variational parameter is adapted by using the ultrarelativistic limit of the period as a boundary condition, which determines the values of additional free parameters introduced. General approximations for the potentials $\frac{x^2}{2}+\frac{x^{2m}}{2m}$, $\sum_{n=1}^m \frac{x^{2m}}{2m}$ and $\frac{x^{2m}}{2m}$ prove to be accurate over the whole solution domain and even for large values of $m$.
\end{abstract}


\maketitle
\section{Introduction}
Interest in the relativistic harmonic oscillator has either focused on its mathematical and physical characteristics \cite{maccoll, moreau, harvey, struble, kim, huang, hutten} or on analytic approaches of solving its non-linear equation of motion, given non-dimensionally as
\begin{align}
\ddot{x}+x(1-\dot{x}^2)^{\frac{3}{2}}=0. \label{harmonic}
\end{align}
The latter is motivated by the general significance of the study of strongly non-linear oscillators, particularly in developing efficient approximation methods to treat them \cite{nayfeh,he,pelster}. 

Recently, the non-perturbative methods of harmonic balance \cite{mickens1,belendez1,belendez2} and homotopy \cite{belendez25,belendez3} have been succesfully applied to give analytic approximations of the oscillation period and periodic solution of Eq.\ref{harmonic}. Unlike traditional perturbation theory, these asymptotic techniques do not rely on any linear term or physically small parameter. Hence their results are unrestricted by a radius of convergence. For both methods, the approximations had minimal error over the entire domain of the particle's maximum velocity, $\dot{x}_{max} \in [0,1)$ \footnote{The regimes of this domain describe low energy oscillations in the region $\dot{x}_{max} << 1$, where Eq.\ref{harmonic} reduces to the linear oscillator. While in the limit $\dot{x}_{max} \to 1$, the oscillator is energetic enough to reach relativistic speeds, which describes a highly non-linear system.}. However, the applicability of these approaches has thus far been restricted to the linear potential.

In this work we contribute a novel method for computing analytic approximations of the period of strongly anharmonic oscillations described by
\begin{align}
\ddot{x}+\frac{d\p}{dx}(1-\dot{x}^2)^{\frac{3}{2}}=0, \label{coordtime}
\end{align}
for an even polynomial potential $\px$ of general form. We adopt a non-perturbative approach that has been developed and applied by Amore et al. in computing periods of non-linear oscillators \cite{amore1,amore2} and formally equivalent systems such as the deflection angle in gravitational lensing \cite{amore3}. The method operates by converting the relevant integral into a series carrying a variational parameter $\w$, which can be chosen using the Principle of Minimal Sensitivity (PMS) \cite{stevenson}. In our case the standard procedure for choosing $\w$ is modified by treating the existing harmonics as free parameters. These are then assigned values by imposing the ultrarelativistic limit of the period as a boundary condition. Although the necessary computations are simple, the resulting approximations are compact and accurate over the whole solution domain.

\section{Method}
As applied to the relativistic harmonic oscillator \cite{mickens1}, the qualitative analysis of ordinary differential equations detailed in Ref.\cite{mickens2} can be used to show that the solution to Eq.\ref{coordtime} is indeed periodic. That is, the phase space trajectories given by
\begin{align}
\frac{dy}{dx}+\frac{d\p}{dx} \frac{(1-y^2)^{\frac{3}{2}}}{y}=0, \label{phase}
\end{align}
where $y=\dot{x}$, are closed and have the strip structure:
\begin{align}
-\infty < x < +\infty \ \ \ \text{and} \ \ \ -1<y<+1.
\end{align}

The period of this system can be derived from the generalized energy-momentum relation
\begin{align}
E=\sqrt{p^2+1}+\px, \label{energymom}
\end{align}
where $E$ is the total energy and $p$ is the relativistic momentum. In all our equations the rest mass and speed of light are set to unity. 

By noting that the total energy is the sum of the potential energy at the amplitude $A$ and the rest mass, i.e. $E=\pa+1$, it can be shown from Eq.\ref{energymom} that
\begin{align}
\pa-\px+1=(1-y^2)^{-\frac{1}{2}} \label{rel1},
\end{align}
which leads to the following expression for the period:
\begin{align}
T=2\int^A_{-A}dx\frac{\pa-\px+1}{\sqrt{[\pa-\px][\pa-\px+2]}}. \label{coordperiod}
\end{align}

Since the period is a function of the oscillation amplitude, it would be useful to express the physical regimes described by $\dot{x}_{max}$ in terms of this parameter. Setting $x=0$ in Eq.\ref{rel1} gives
\begin{align}
\dot{x}_{max}=\frac{\sqrt{\pa[2+\pa]}}{1+\pa}.
\end{align}
This relation maps the domain $\dot{x}_{max} \in [0,1)$ unto $A \in [0,\infty)$.

For an even potential $\px$, the period from Eq.\ref{coordperiod} can be written as
\begin{align}
T=2\int^A_{-A}\frac{dx}{\sqrt{\Rx[A^2-x^2]}}. \label{coordperiod2}
\end{align}
We approximate $\Rx(A^2-x^2)$ with the solvable harmonic potential $\w^2(A^2-x^2)$, in a manner that is essentially a Linear Delta Expansion \cite{amore3,amore4}:
\begin{align}
T=2\int^A_{-A}\frac{dx}{\sqrt{[A^2-x^2][\w^2+\delta(\Rx-\w^2)]}}. \label{lde}
\end{align}
In the above equation, when $\delta=0$ we have the case of simple harmonic motion, while for $\delta=1$ the system reduces to the original potential of Eq.\ref{coordperiod2}.

With a change of variable $x=A\cx$, the above equation becomes
\begin{align}
T=\frac{2}{\w}\int_0^{\pi} \frac{d\theta}{\sqrt{1+\delta \Delta\bm{(}\theta,\w\bm{)}}}, \label{period2}
\end{align}
where the substitution $\Delta\bm{(}\theta,\w\bm{)}=-1+\frac{\Rt}{\w^2}$ has been made. A binomial expansion of the integrand in powers of $\Delta\bm{(}\theta,\w\bm{)}$ then gives

\begin{align} 
T=\frac{2}{\w}\sum_{n=0}^{\infty} \tiny{\bigr{(} \hspace{-0.15cm} \begin{array}{cc}n&\\-1/2\end{array} \hspace{-0.26cm}\bigr{)}} \int_0^{\pi} \D^n\bm{(}\t,\w \bm{)}  d\t, \label{series}
\end{align}
where we have set $\delta=1$.

Evaluating the above series to a finite order will leave an explicit $\w$-dependence that should be eliminated. Indeed this is the basis of the PMS, requiring that $\frac{\partial T}{\partial \w}=0$. As shown in Ref.\cite{amore2}, the condition is equivalent to choosing $\w$ such that
\begin{align}
\int_0^{\pi} \D^n\bm{(}\t,\w\bm{)}d\t=0 \label{PMS}
\end{align}
for an nth order optimization. To first order, the above criterion is equivalently
\begin{align}
\w=\sqrt{\left\langle \Rt \right\rangle} \label{PMS2},
\end{align}
where $\left\langle \Rt \right\rangle$ denotes the mean value of $\Rt$.

For classical oscillators Eq.\ref{PMS2} is implemented easily, yielding remarkable results \cite{amore2}. In the case of relativistic oscillators however, the expressions not only become cumbersome but are divergent at large oscillation amplitudes even with higher order computations. For example, we find that the resulting $\w$ computed for the relativistic quartic oscillator approaches unity as $A \to \infty$. This marks an ineffective variation since the choice $\w=1$ gives the exact untreated series. Hence for any order of calculation, the error of approximation increases rapidly beyond $A\approx1$ ($\dot{x}_{max} \approx 0.82$).

The alternative criterion given by Eq.36 of Ref.\cite{amore2} is also found to be problematic because the resulting $\w$ goes to infinity as $A \to \infty$. Since all terms in Eq.\ref{series} are proportional to powers of $\frac{1}{\w}$, then any finite order approximation incorrectly goes to zero in the ultrarelativistic limit $A \to \infty$.

The correct large amplitude behavior can be verified by noting from Eq.\ref{rel1} that $\pa-\px>>1$ in this regime. Hence, Eq.\ref{coordperiod} simplifies to what is expected for a photon bouncing in a box of width $2A$:
\begin{align}
T \sim 4A. \label{uperiod}
\end{align}
The preceding limit will serve as the asymptotic boundary condition that determines our variational parameter.

We begin constructing our solution by noting that the zeroth order truncation of the series in Eq.\ref{series} is
\begin{align}
T=\frac{2\pi}{\w}. \label{approx}
\end{align}
We then assume the form of $\w$ that directly satisfies Eq.\ref{PMS}, namely
\begin{align}
\w=\sqrt{\Rt}. \label{PMS3}
\end{align}
In the above equation $\w$ is still dependent on the variable of integration $\t$ since harmonics of the form $\cos^{2k}\t$ are present. To remove this, we treat the harmonics as free parameters, denoting $\cos^{2k}\t$ as $\lambda_{2k}$. 

Next, we look at the limit of $\w$ as $A \to \infty$. It can be shown that for an even potential with leading order $x^{2m}$, the large amplitude behavior is
\begin{align}
\w \sim \frac{1}{A}\sqrt{\frac{\sum_{n=0}^{m-1} (\lambda_{2n}-\lambda_{2(m+n)})}{1-2\lambda_{2m}+\lambda_{4m}}}. \label{wlarge1}
\end{align}
Substituting Eqs.\ref{uperiod} and \ref{wlarge1} in the limit $A \to \infty$ of Eq.\ref{approx} will yield a relation between all orders of free parameters that exist in $R$. This relation can be used recursively to provide $m$ equations. However for a non-linear potential, the system of equations is under-determined since there are a total of $2m$ free parameters. We remedy this problem by factoring out $1-\lambda_{2m}$ from the numerator and denominator in Eq.\ref{wlarge1}, giving us
\begin{align}
\w \sim \frac{1}{A}\sqrt{\frac{\sum_{n=0}^{m-1} \lambda_{2n}}{1-\lambda_{2m}}}. \label{wlarge2}
\end{align}
Thus the large amplitude limit of Eq.\ref{approx} yields the relation
\begin{align}
\lambda_{2m}=1-\frac{4}{\pi^2}\sum_{n=0}^{m-1} \lambda_{2n}. \label{harmonics1}
\end{align}
Under the ansatz of setting the lowest order free parameter to be its trigonometric average, i.e. $\lambda_2=\frac{1}{2}$, the $m=2$ case of the above equation determines $\lambda_4$. These values are then used in the $m=3$ equation to solve for $\lambda_6$ and so on, effectively building the whole set of $m$ free parameters from the bottom up. This iterative computation yields that for integers $n\geq2$ the free parameters are given by
\begin{align}
\lambda_{2n}=\frac{\pi^2-6}{\pi^2-4} \left(\frac{\pi^2-4}{\pi^2}\right)^{n-1}. \label{free}
\end{align}
We use these values in $R$ and determine $\w$ by Eq.\ref{PMS3}. Then by Eq.\ref{approx} the resulting approximation is
\begin{align}
T=\frac{2 \pi}{\sqrt{R\bm{(}\lambda_{2},\lambda_{4},...,\lambda_{2m}\bm{)}}}. \label{approx2}
\end{align}

\begin{figure}[tb]
\begin{center}
   \includegraphics[bb = 0 0 469 341, width=0.7\linewidth,clip]{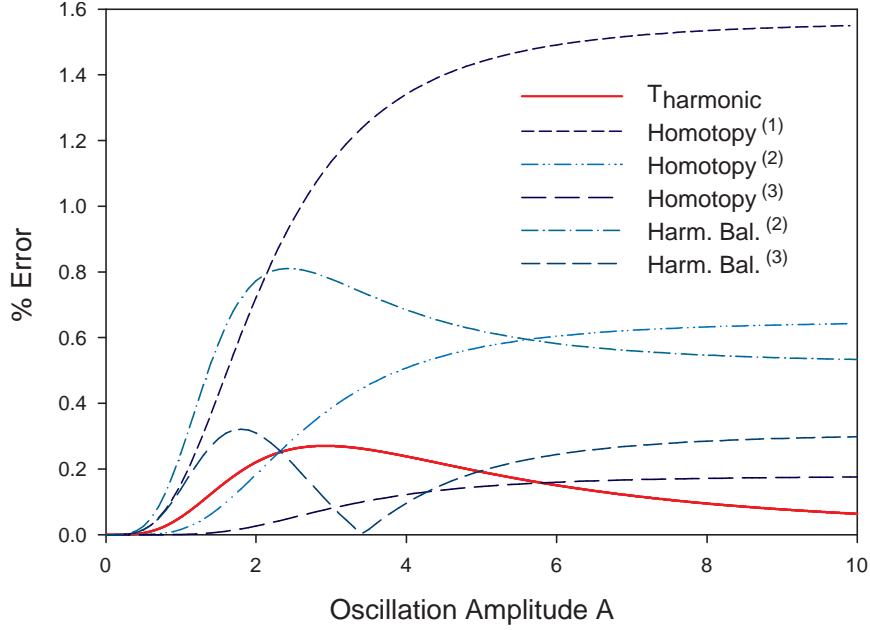}
\end{center}
\caption{(Color online) Relative error of Eq.\ref{harmperiod} with respect to the exact result given by Eq.\ref{exact}. Shown for comparison are the relative error of the second and third order computations of the harmonic balance method \cite{belendez1,belendez2} and the first to third order computations of the homotopy method \cite{belendez3}.}\label{fig1}
\end{figure}

\section{Applications}
As a first example we apply the method to the linear potential. Plugging in $\px=\frac{x^2}{2}$ in Eq.\ref{coordperiod} and following the general derivation previously outlined yields
\begin{align}
R=\frac{4+A^2(1-\lambda_2)}{4+4A^2(1-\lambda_2)+A^4(1-2\lambda_2+\lambda_4)}. \label{harmR}
\end{align}
In calculating the free parameters present in $R$ we utilize the relation given by Eq.\ref{wlarge1} since for $m=1$ the system is not under-determined. Accordingly, the ultrarelativistic limit of Eq.\ref{approx} gives the relation
\begin{align}
\frac{2}{\pi}=\sqrt{\frac{1-2\lambda_2+\lambda_4}{1-\lambda_2}}.
\end{align}
By using $\lambda_2=\frac{1}{2}$ in the above equation, we determine $\lambda_4=\frac{2}{\pi^2}$. We then substitute these values into Eq.\ref{harmR} and use Eq.\ref{approx2}, giving us the compact expression for the period
\begin{align}
T_{H}=\frac{4 \sqrt{A^4+\left(2+A^2\right) \pi ^2}}{\sqrt{8+A^2}}. \label{harmperiod}
\end{align}

We compare this with the exact period given by
\begin{align}
T_{exact}=\frac{4 \left(4+A^2\right) \text{E}\left[\frac{A^2}{4+A^2}\right]-8 \text{K}\left[\frac{A^2}{4+A^2}\right]}{\sqrt{4+A^2}}, \label{exact}
\end{align}
where $\text{E}[\cdot]$ and $\text{K}[\cdot]$ are the Elliptic Integrals of the First and Second kind, respectively. Figure \ref{fig1} shows the relative error of Eq.\ref{harmperiod} along with those of the results of harmonic balance \cite{belendez1,belendez2} and homotopy \cite{belendez3}. Despite the simplicity of our calculation, the third order approximations of both these methods are only more accurate by $0.1-0.2\%$ over a finite range of $A$. Our solution is the only approximation whose error goes to zero at the ultrarelativistic limit. An advantage is also apparent in the resulting expression itself, that it is compact and mathematically elementary, while those of harmonic balance and homotopy are lengthy and even require the evaluation of Elliptic Integrals.

The most note-worthy point of our method is its extensive applicability. For our second example we calculate a general approximation for the periods of relativistic oscillators with the potential $\px=\frac{x^2}{2}+\frac{x^{2m}}{2m}$. As we have done previously, we obtain the function $R$ by following the derivation starting from Eq.\ref{coordperiod}. The result is
\begin{align}
R=(m+A^{2m-2} \sum_{n=0}^{m-1} \lambda_{2n})\frac{(4m+Z_1)}{(2m+Z_1)^2}, \label{genR}
\end{align}
where
\begin{align}
Z_1=mA^2(1-\lambda_2)+A^{2m}(1-\lambda_{2m}). \label{z1}
\end{align}
Note that this expression has been factored in a manner compatible with Eq.\ref{wlarge2} so that the existing free parameters are defined by Eq.\ref{free}. Then the approximation simply follows from Eq.\ref{approx2}. Explicitly, the result is
\begin{align}
T_a=\frac{2\pi (2m+Z_1)}{\sqrt{(4m+Z_1)\{m+A^{2m-2} [\frac{3}{2}+\frac{\pi^2-6}{4}(1-(\frac{\pi^2-4}{\pi^2})^{m-2})]\}}}, \label{genT}
\end{align}
where we have evaluated the geometric sum appearing in Eq.\ref{genR}. To give specific examples of this general result, the approximations for the periods of the relativistic quartic ($m=2$) and sextic ($m=3$) oscillators are respectively
\begin{align}
T_{quartic}=\frac{2 \sqrt{2} [6 A^4+(4+A^2) \pi ^2]}{\sqrt{(4+3 A^2)[6 A^4+(8+A^2) \pi ^2]}} \label{duffperiod}
\end{align}
and
\begin{align}
T_{sextic}=\frac{24 \pi ^4+6 A^2 \pi ^4+A^6 \left(40 \pi ^2-96\right)}{\sqrt{[6 \pi ^2+A^4 (5 \pi ^2-12)][24 \pi ^4+3 A^2 \pi ^4+A^6 (20 \pi ^2-48)]}}. \label{qsperiod}
\end{align}

\begin{figure}[tb]
\begin{center}
   \includegraphics[bb = 0 0 583 421, width=0.7\linewidth,clip]{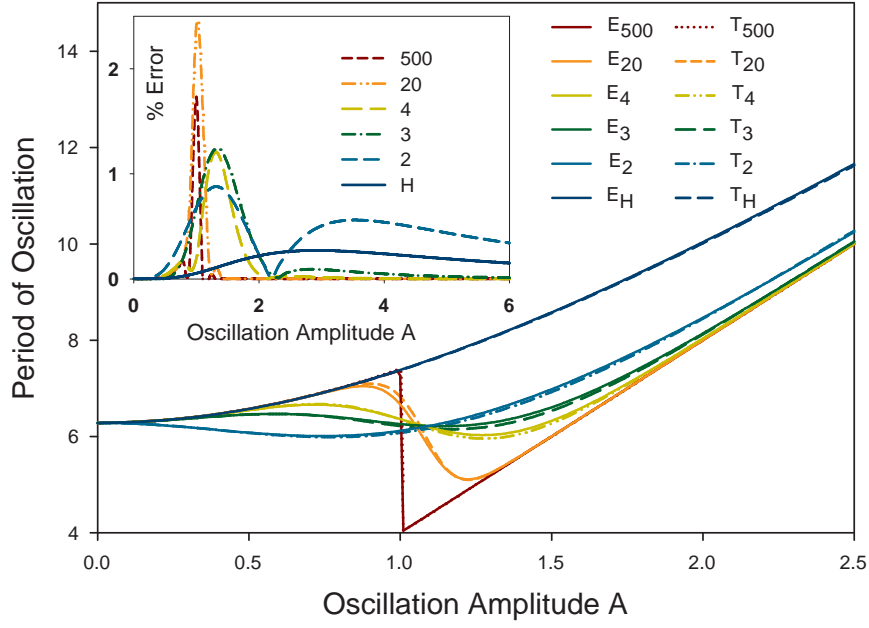}
\end{center}
\caption{(Color online) Numerically integrated exact results (approximations) for the cases $m=2,3,4,20,500$ and the harmonic potential are denoted by $E_m$ ($T_m$) and $E_H$ ($T_H$), respectively. The inset shows relative errors labeled by the respective subscripts.}\label{fig2}
\end{figure}

The approximations for the cases $m=2,3,4,20,500$ and the harmonic potential are fitted with the numerically integrated exact results in Figure \ref{fig2}. The relative errors shown in the inset are remarkably low even for the highly non-linear case of $m=500$. For such large values of $m$ the period exhibits a sharp jump from what is characteristically harmonic at amplitudes $A<1$ to the ultrarelativistic limit for $A\geq1$.

As a third example, we derive a general approximation for the periods of relativistic oscillators with the potential $\px=\sum_{n=1}^m \frac{x^{2m}}{2m}$. Again, the crux of the calculation is deriving a general form of $R$ and factoring it compatible with Eq.\ref{wlarge2} so that the free parameters are determined by Eq.\ref{free}. Then by using Eq.\ref{approx2}, the following result is obtained
\begin{align}
T_b=\frac{2\pi (1+Z_2)}{\sqrt{(2+Z_2)\{\frac{1}{2}+\sum_{n=2}^m\frac{A^{2n-2}}{2n}[\frac{3}{2}+\frac{\pi^2-6}{4}(1-(\frac{\pi^2-4}{\pi^2})^{n-2})]\}}}, \label{genT2}
\end{align}
where
\begin{align}
Z_2=\sum_{n=1}^m \frac{A^{2n-2}}{2n}(1-\lambda_{2n}). \label{z2}
\end{align}

\begin{figure}[tb]
\begin{center}
   \includegraphics[bb = 0 0 417 302, width=0.7\linewidth,clip]{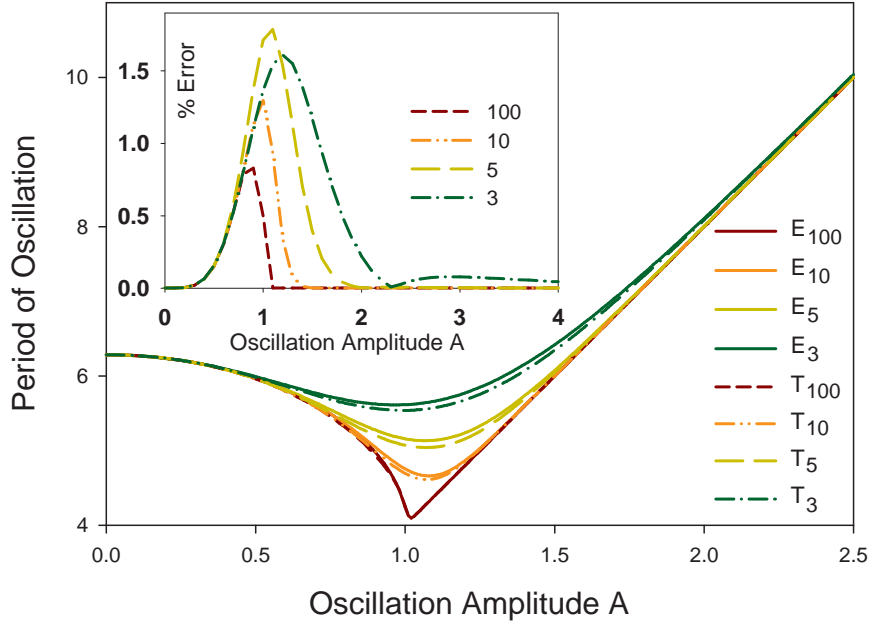}
\end{center}
\caption{(Color online) Numerically integrated exact results and approximations for the cases $m=3,5,10$ and $100$ are denoted by $E_m$ and $T_m$, respectively. The inset shows relative errors labeled by the respective subscripts.}\label{fig3}
\end{figure}

For $m=2$ this reduces to Eq.\ref{duffperiod}, the period of a relativistic quartic oscillator. For $m=3$, it provides the following approximation to the period of a relativistic particle oscillating under the potential $\px=\frac{x^2}{2}+\frac{x^4}{4}+\frac{x^6}{6}$
\begin{align}
T_{2,4,6}=\frac{2 \sqrt{2}[-48 A^6+2 A^4 (9+10 A^2) \pi ^2+3 (4+A^2) \pi ^4]}{\sqrt{[-24 A^4+(12+9 A^2+10 A^4) \pi ^2][-48 A^6+2 A^4 (9+10 A^2) \pi ^2+3 (8+A^2) \pi ^4 ]}}. \label{period246}
\end{align}

Figure \ref{fig3} shows the approximations for the cases $m=3,5,10$ and $100$, fitted with the numerically integrated exact results. Notably, the inset shows low relative errors whose maximum do not increase with increasing $m$.

\begin{figure}[tb]
\begin{center}
   \includegraphics[bb = 0 0 417 305, width=0.7\linewidth,clip]{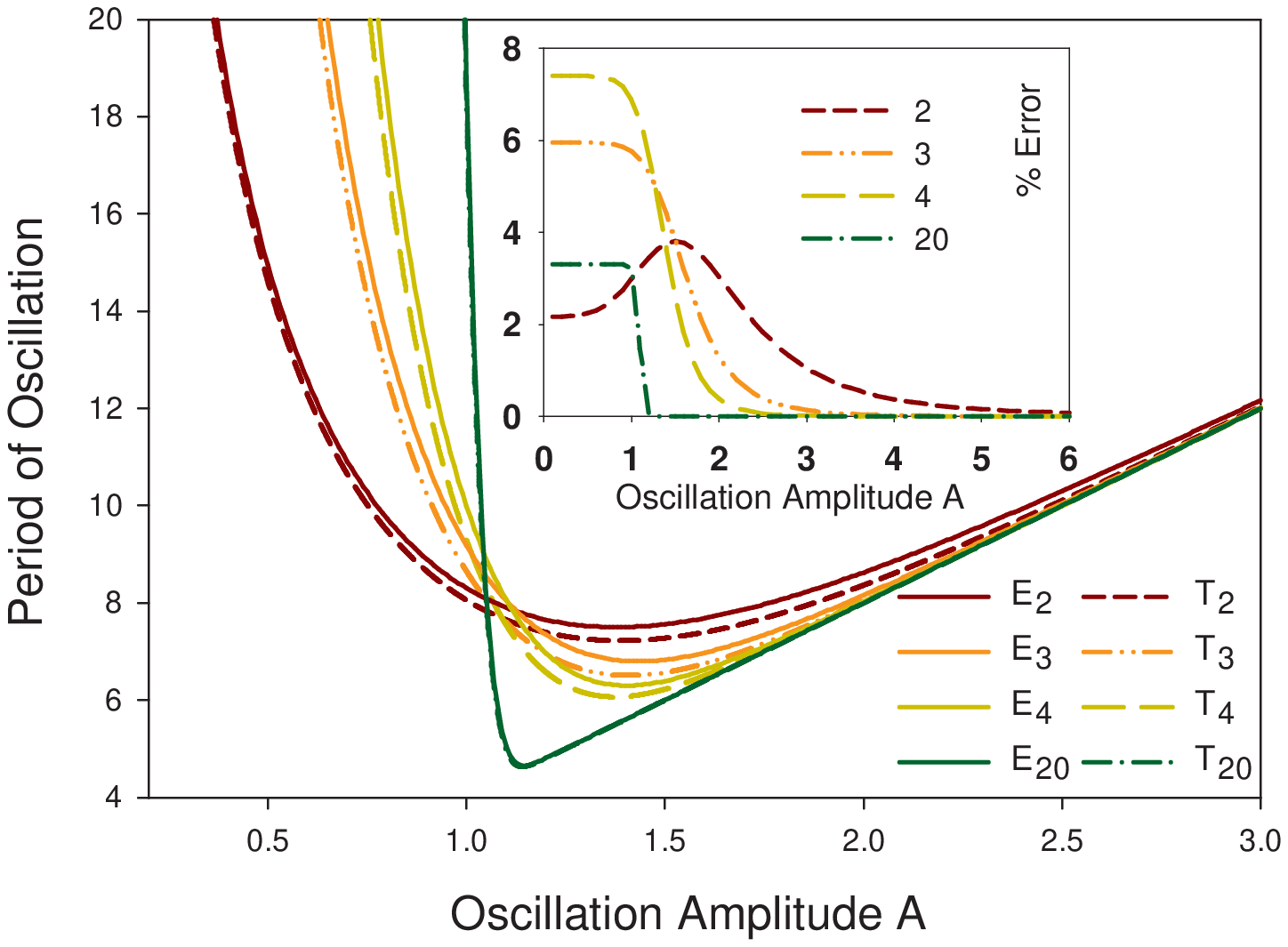}
\end{center}
\caption{(Color online) Numerically integrated exact results and approximations for the cases $m=2,3,4$ and $20$ are denoted by $E_m$ and $T_m$, respectively. The inset shows relative errors labeled by the respective subscripts.}\label{fig4}
\end{figure}

Lastly, we treat relativistic oscillators with the intrinsically non-linear potential $\frac{x^{2m}}{2m}$. Using the same procedure of deriving a general expression for $R$ and using Eqs.\ref{free} and \ref{approx2}, the result is
\begin{align}
T_c=\frac{2\pi [2m+A^{2m}(1-\lambda_{2m})]}{A^{m-1}\sqrt{[4m+A^{2m}(1-\lambda_{2m})] \{\frac{3}{2}+\frac{\pi^2-6}{4}[1-(\frac{\pi^2-4}{\pi^2})^{m-2} ]\} }} . \label{genT3}
\end{align}
For example, setting $m=2$ in the equation above gives the following approximation for the period of a relativistic oscillator with potential $\frac{x^4}{4}$
\begin{align}
T_{x^4}=\frac{12 A^4 \pi +8 \pi ^3}{\sqrt{9 A^6 \pi ^2+12 A^2 \pi ^4}}.
\end{align}

The approximations for the cases $m=2,3,4$ and $20$ and their numerically integrated exact results are shown in Figure \ref{fig4}. The method is able to capture the infinities at both regimes of the oscillation period. 

\section{Conclusions}\label{summary}
Our modification of a PMS-based non-perturbative method offers an economical procedure of calculating compact and accurate analytic expressions for the period of an oscillating relativistic particle. It is able to attack even polynomial potentials of general form, including those that are intrinsically non-linear. The mechanics of the method itself guarantees the accuracy of the solution even at the ultrarelativistic limit.

J.~P.~E.~acknowledges support from a National Institute of Physics Faculty Grant.

\end{document}